\documentstyle[10pt,epsf,epsfig,dp_delphititle,oldlfont]{dp_delphi}
\makeindex
\def\DpPaperGroup{EP}
\def\DpPaperRef{2002-023}
\def\DpDate{12 March 2002}
\def\DpAuthors{DELPHI Collaboration}
\def\DpSubmit{(Accepted by Phys.Lett.B)}
\def\DpTitle{{\boldmath
     Rapidity-Alignment and $\mrm{p}_T$\ Compensation \\
       of Particle Pairs in Hadronic \zzero~Decays}}
\def\DpComment{ }
\def\DpEMail{}

\newcommand{\mrm} {\mathrm}
\newcommand{\bfm} {\boldmath}

\newcommand{\zzero}{${{\mrm{Z}^0}}$}
\newcommand{\into}{{\ifmmode \rightarrow \else $\rightarrow$\fi}}
\newcommand{\epem}{{\ifmmode e^+e^- \else $e^+e^-$\fi}}
\newcommand{\zz}{{\ifmmode Z^0~ \else $Z^0~$\fi}}
\newcommand{\sqs}{{\ifmmode \sqrt s~ \else $\sqrt s~$\fi}}
\newcommand{\de}{{\ifmmode ^{\circ} \else ${^\circ}$\fi}}
\newcommand{\q}{{\ifmmode q~ \else $q~$\fi}}
\newcommand{\qb}{{\ifmmode \bar{q}~ \else $\bar{q}~$\fi}}
\newcommand{\sx}{{\ifmmode s~ \else $s~$\fi}}
\newcommand{\sxb}{{\ifmmode \bar{s}~ \else $\bar{s}~$\fi}}
\newcommand{\cxb}{{\ifmmode \bar{c}~ \else $\bar{c}~$\fi}}
\newcommand{\sxx}{{\ifmmode s \else $s$\fi}}
\newcommand{\sxbx}{{\ifmmode \bar{s} \else $\bar{s}$\fi}}
\newcommand{\Kp}{{\ifmmode K^{+}~ \else $K^{+}~$\fi}}
\newcommand{\Kpx}{{\ifmmode K^{+} \else $K^{+}$\fi}}
\newcommand{\Km}{{\ifmmode K^{-}~ \else $K^{-}~$\fi}}
\newcommand{\Kmx}{{\ifmmode K^{-} \else $K^{-}$\fi}}
\newcommand{\Dz}{{\ifmmode D^{0}~ \else $D^{0}~$\fi}}
\newcommand{\Dp}{{\ifmmode D^{+}~ \else $D^{+}~$\fi}}
\newcommand{\p}{{\ifmmode p~ \else $p~$\fi}}
\newcommand{\px}{{\ifmmode p \else $p$\fi}}
\newcommand{\bp}{{\ifmmode \bar{p}~ \else $\bar{p}~$\fi}}
\newcommand{\bpx}{{\ifmmode \bar{p} \else $\bar{p}$\fi}}

\newcommand{\qqb}{{\ifmmode q\,\bar{q}~ \else $q\,\bar{q}~$\fi}}
\newcommand{\ssb}{{\ifmmode s\,\bar{s}~ \else $s\,\bar{s}~$\fi}}
\newcommand{\ccb}{{\ifmmode c\,\bar{c}~ \else $c\,\bar{c}~$\fi}}
\newcommand{\bbb}{{\ifmmode b\,\bar{b}~ \else $b\,\bar{b}~$\fi}}
\newcommand{\pipi}{{\ifmmode \pi^+\,\pi^-~ \else $\pi^+\,\pi^-~$\fi}}
\newcommand{\pipix}{{\ifmmode \pi^+\,\pi^- \else $\pi^+\,\pi^-$\fi}}
\newcommand{\KK}{{\ifmmode K^+K^-~ \else $K^+K^-~$\fi}}
\newcommand{\KKx}{{\ifmmode K^+K^- \else $K^+K^-$\fi}}
\newcommand{\pp}{{\ifmmode p\,\bar{p}~ \else $p\,\bar{p}~$\fi}}
\newcommand{\ppx}{{\ifmmode p\,\bar{p} \else $p\,\bar{p}$\fi}}
\newcommand{\dy}{{\ifmmode {\mit\Delta y}~ \else ${\mit\Delta y}~$\fi}}
\newcommand{\dyx}{{\ifmmode {\mit\Delta y} \else ${\mit\Delta y}$\fi}}
\newcommand{\PPP}{{\ifmmode \cal{P}~ \else $\cal{P}~$\fi}}
\newcommand{\PPPx}{{\ifmmode \cal{P} \else $\cal{P}$\fi}}
\newcommand{\PPPdy}{{$\PPPx(\dyx)~$}}

\newcommand{\pT}{{$p_T~$}}
\newcommand{\pTx}{{$p_T$}}
\newcommand{\dpT}{{\ifmmode \Delta\pT~ \else $\Delta\pT~$\fi}}
\newcommand{\dpTx}{{\ifmmode \Delta\pT \else $\Delta\pT$\fi}}
\newcommand{\Nin}{{\ifmmode N_{in}~ \else $N_{in}~$\fi}}
\newcommand{\Ninx}{{\ifmmode N_{in} \else $N_{in}$\fi}}
\newcommand{\Nout}{{\ifmmode N_{out}~ \else $N_{out}~$\fi}}
\newcommand{\Noutx}{{\ifmmode N_{out} \else $N_{out}$\fi}}
\newcommand{\dphi}{{\ifmmode {\mit\Delta}\phi~ \else ${\mit\Delta}\phi~$\fi}}
\newcommand{\dphix}{{\ifmmode {\mit\Delta}\phi \else ${\mit\Delta}\phi $\fi}}
\begin{document}
\makeatletter
\newcount\@tempcntc
\def\@citex[#1]#2{\if@filesw\immediate\write\@auxout{\string\citation{#2}}\fi
  \@tempcnta\z@\@tempcntb\m@ne\def\@citea{}\@cite{\@for\@citeb:=#2\do
    {\@ifundefined
       {b@\@citeb}{\@citeo\@tempcntb\m@ne\@citea\def\@citea{,}{\bf ?}\@warning
       {Citation `\@citeb' on page \thepage \space undefined}}%
    {\setbox\z@\hbox{\global\@tempcntc0\csname b@\@citeb\endcsname\relax}%
     \ifnum\@tempcntc=\z@ \@citeo\@tempcntb\m@ne
       \@citea\def\@citea{,}\hbox{\csname b@\@citeb\endcsname}%
     \else
      \advance\@tempcntb\@ne
      \ifnum\@tempcntb=\@tempcntc
      \else\advance\@tempcntb\m@ne\@citeo
      \@tempcnta\@tempcntc\@tempcntb\@tempcntc\fi\fi}}\@citeo}{#1}}
\def\@citeo{\ifnum\@tempcnta>\@tempcntb\else\@citea\def\@citea{,}%
  \ifnum\@tempcnta=\@tempcntb\the\@tempcnta\else
   {\advance\@tempcnta\@ne\ifnum\@tempcnta=\@tempcntb \else \def\@citea{--}\fi
    \advance\@tempcnta\m@ne\the\@tempcnta\@citea\the\@tempcntb}\fi\fi}
 
\makeatother
\begin{titlepage}
\pagenumbering{roman}
\CERNpreprint{\DpPaperGroup}{\DpPaperRef} 
\date{{\small\DpDate}} 
\title{\DpTitle} 
\address{\DpAuthors} 
\begin{shortabs} 
\noindent
Observation is made of rapidity-alignment of \KK and \pp pairs 
which results from their asymmetric orientation in rapidity, with respect
to the direction from primary quark to antiquark.
The \KK and \pp data are consistent with predictions from the fragmentation 
string model.
However, the \pp data strongly disagree with the conventional implementation
of the cluster model.
The non-perturbative process of `gluon splitting to diquarks' has to be
incorporated into the cluster model for it to agree with the data.
Local conservation of \pT between particles nearby in rapidity (i.e., \pT
compensation) is analysed with respect to the thrust direction for \pipix,
\KKx, and \pp pairs.
In this case, the string model provides fair agreement with the data.
The cluster model is incompatible with the data for all three particle pairs.
The model with its central premiss of isotropically-decaying clusters
predicts a \pT correlation not seen in the data.

\end{shortabs}
\vfill
\begin{center}
\DpSubmit \ \\ 
\DpComment \ \\
\DpEMail \ \\
\end{center}
\vfill
\clearpage
\headsep 10.0pt
\addtolength{\textheight}{10mm}
\addtolength{\footskip}{-5mm}
\begingroup
%
\newcommand{\DpName}[2]{\hbox{#1$^{\ref{#2}}$},\hfill}
\newcommand{\DpNameTwo}[3]{\hbox{#1$^{\ref{#2},\ref{#3}}$},\hfill}
\newcommand{\DpNameThree}[4]{\hbox{#1$^{\ref{#2},\ref{#3},\ref{#4}}$},\hfill}
\newskip\Bigfill \Bigfill = 0pt plus 1000fill
\newcommand{\DpNameLast}[2]{\hbox{#1$^{\ref{#2}}$}\hspace{\Bigfill}}
%
\footnotesize
\noindent
\DpName{J.Abdallah}{LPNHE}
\DpName{P.Abreu}{LIP}
\DpName{W.Adam}{VIENNA}
\DpName{P.Adzic}{DEMOKRITOS}
\DpName{T.Albrecht}{KARLSRUHE}
\DpName{T.Alderweireld}{AIM}
\DpName{R.Alemany-Fernandez}{CERN}
\DpName{T.Allmendinger}{KARLSRUHE}
\DpName{P.P.Allport}{LIVERPOOL}
\DpName{S.Almehed}{LUND}
% ...
\normalsize
\endgroup
\titlefoot{Department of Physics and Astronomy, Iowa State
     University, Ames IA 50011-3160, USA
    \label{AMES}}
\titlefoot{Physics Department, Universiteit Antwerpen,
     Universiteitsplein 1, B-2610 Antwerpen, Belgium \\
     \indent~~and IIHE, ULB-VUB,
     Pleinlaan 2, B-1050 Brussels, Belgium \\
     \indent~~and Facult\'e des Sciences,
     Univ. de l'Etat Mons, Av. Maistriau 19, B-7000 Mons, Belgium
    \label{AIM}}
\titlefoot{Physics Laboratory, University of Athens, Solonos Str.
     104, GR-10680 Athens, Greece
    \label{ATHENS}}
\titlefoot{Department of Physics, University of Bergen,
     All\'egaten 55, NO-5007 Bergen, Norway
    \label{BERGEN}}
\titlefoot{Dipartimento di Fisica, Universit\`a di Bologna and INFN,
     Via Irnerio 46, IT-40126 Bologna, Italy
    \label{BOLOGNA}}
\titlefoot{Centro Brasileiro de Pesquisas F\'{\i}sicas, rua Xavier Sigaud 150,
     BR-22290 Rio de Janeiro, Brazil \\
     \indent~~and Depto. de F\'{\i}sica, Pont. Univ. Cat\'olica,
     C.P. 38071 BR-22453 Rio de Janeiro, Brazil \\
     \indent~~and Inst. de F\'{\i}sica, Univ. Estadual do Rio de Janeiro,
     rua S\~{a}o Francisco Xavier 524, Rio de Janeiro, Brazil
    \label{BRASIL}}
\titlefoot{Coll\`ege de France, Lab. de Physique Corpusculaire, IN2P3-CNRS,
     FR-75231 Paris Cedex 05, France
    \label{CDF}}
\titlefoot{CERN, CH-1211 Geneva 23, Switzerland
    \label{CERN}}
\titlefoot{Institut de Recherches Subatomiques, IN2P3 - CNRS/ULP - BP20,
     FR-67037 Strasbourg Cedex, France
    \label{CRN}}
\titlefoot{Now at DESY-Zeuthen, Platanenallee 6, D-15735 Zeuthen, Germany
    \label{DESY}}
\titlefoot{Institute of Nuclear Physics, N.C.S.R. Demokritos,
     P.O. Box 60228, GR-15310 Athens, Greece
    \label{DEMOKRITOS}}
\titlefoot{Dipartimento di Fisica, Universit\`a di Genova and INFN,
     Via Dodecaneso 33, IT-16146 Genova, Italy
    \label{GENOVA}}
\titlefoot{Institut des Sciences Nucl\'eaires, IN2P3-CNRS, Universit\'e
     de Grenoble 1, FR-38026 Grenoble Cedex, France
    \label{GRENOBLE}}
\titlefoot{Helsinki Institute of Physics, HIP,
     P.O. Box 9, FI-00014 Helsinki, Finland
    \label{HELSINKI}}
\titlefoot{Joint Institute for Nuclear Research, Dubna, Head Post
     Office, P.O. Box 79, RU-101 000 Moscow, Russian Federation
    \label{JINR}}
\titlefoot{Institut f\"ur Experimentelle Kernphysik,
     Universit\"at Karlsruhe, Postfach 6980, DE-76128 Karlsruhe,
     Germany
    \label{KARLSRUHE}}
\titlefoot{Institute of Nuclear Physics,Ul. Kawiory 26a,
     PL-30055 Krakow, Poland
    \label{KRAKOW1}}
\titlefoot{Faculty of Physics and Nuclear Techniques, University of Mining
     and Metallurgy, PL-30055 Krakow, Poland
    \label{KRAKOW2}}
\titlefoot{Universit\'e de Paris-Sud, Lab. de l'Acc\'el\'erateur
     Lin\'eaire, IN2P3-CNRS, B\^{a}t. 200, FR-91405 Orsay Cedex, France
    \label{LAL}}
\titlefoot{School of Physics and Chemistry, University of Landcaster,
     Lancaster LA1 4YB, UK
    \label{LANCASTER}}
\titlefoot{LIP, IST, FCUL - Av. Elias Garcia, 14-$1^{o}$,
     PT-1000 Lisboa Codex, Portugal
    \label{LIP}}
\titlefoot{Department of Physics, University of Liverpool, P.O.
     Box 147, Liverpool L69 3BX, UK
    \label{LIVERPOOL}}
\titlefoot{LPNHE, IN2P3-CNRS, Univ.~Paris VI et VII, Tour 33 (RdC),
     4 place Jussieu, FR-75252 Paris Cedex 05, France
    \label{LPNHE}}
\titlefoot{Department of Physics, University of Lund,
     S\"olvegatan 14, SE-223 63 Lund, Sweden
    \label{LUND}}
\titlefoot{Universit\'e Claude Bernard de Lyon, IPNL, IN2P3-CNRS,
     FR-69622 Villeurbanne Cedex, France
    \label{LYON}}
\titlefoot{Dipartimento di Fisica, Universit\`a di Milano and INFN-MILANO,
     Via Celoria 16, IT-20133 Milan, Italy
    \label{MILANO}}
\titlefoot{Dipartimento di Fisica, Univ. di Milano-Bicocca and
     INFN-MILANO, Piazza della Scienza 2, IT-20126 Milan, Italy
    \label{MILANO2}}
\titlefoot{IPNP of MFF, Charles Univ., Areal MFF,
     V Holesovickach 2, CZ-180 00, Praha 8, Czech Republic
    \label{NC}}
\titlefoot{NIKHEF, Postbus 41882, NL-1009 DB
     Amsterdam, The Netherlands
    \label{NIKHEF}}
\titlefoot{National Technical University, Physics Department,
     Zografou Campus, GR-15773 Athens, Greece
    \label{NTU-ATHENS}}
\titlefoot{Physics Department, University of Oslo, Blindern,
     NO-0316 Oslo, Norway
    \label{OSLO}}
\titlefoot{Dpto. Fisica, Univ. Oviedo, Avda. Calvo Sotelo
     s/n, ES-33007 Oviedo, Spain
    \label{OVIEDO}}
\titlefoot{Department of Physics, University of Oxford,
     Keble Road, Oxford OX1 3RH, UK
    \label{OXFORD}}
\titlefoot{Dipartimento di Fisica, Universit\`a di Padova and
     INFN, Via Marzolo 8, IT-35131 Padua, Italy
    \label{PADOVA}}
\titlefoot{Rutherford Appleton Laboratory, Chilton, Didcot
     OX11 OQX, UK
    \label{RAL}}
\titlefoot{Dipartimento di Fisica, Universit\`a di Roma II and
     INFN, Tor Vergata, IT-00173 Rome, Italy
    \label{ROMA2}}
\titlefoot{Dipartimento di Fisica, Universit\`a di Roma III and
     INFN, Via della Vasca Navale 84, IT-00146 Rome, Italy
    \label{ROMA3}}
\titlefoot{DAPNIA/Service de Physique des Particules,
     CEA-Saclay, FR-91191 Gif-sur-Yvette Cedex, France
    \label{SACLAY}}
\titlefoot{Instituto de Fisica de Cantabria (CSIC-UC), Avda.
     los Castros s/n, ES-39006 Santander, Spain
    \label{SANTANDER}}
\titlefoot{Inst. for High Energy Physics, Serpukov
     P.O. Box 35, Protvino, (Moscow Region), Russian Federation
    \label{SERPUKHOV}}
\titlefoot{J. Stefan Institute, Jamova 39, SI-1000 Ljubljana, Slovenia
     and Laboratory for Astroparticle Physics,\\
     \indent~~Nova Gorica Polytechnic, Kostanjeviska 16a, SI-5000 Nova Gorica, Slovenia, \\
     \indent~~and Department of Physics, University of Ljubljana,
     SI-1000 Ljubljana, Slovenia
    \label{SLOVENIJA}}
\titlefoot{Fysikum, Stockholm University,
     Box 6730, SE-113 85 Stockholm, Sweden
    \label{STOCKHOLM}}
\titlefoot{Dipartimento di Fisica Sperimentale, Universit\`a di
     Torino and INFN, Via P. Giuria 1, IT-10125 Turin, Italy
    \label{TORINO}}
\titlefoot{Dipartimento di Fisica, Universit\`a di Trieste and
     INFN, Via A. Valerio 2, IT-34127 Trieste, Italy \\
     \indent~~and Istituto di Fisica, Universit\`a di Udine,
     IT-33100 Udine, Italy
    \label{TU}}
\titlefoot{Univ. Federal do Rio de Janeiro, C.P. 68528
     Cidade Univ., Ilha do Fund\~ao
     BR-21945-970 Rio de Janeiro, Brazil
    \label{UFRJ}}
\titlefoot{Department of Radiation Sciences, University of
     Uppsala, P.O. Box 535, SE-751 21 Uppsala, Sweden
    \label{UPPSALA}}
\titlefoot{IFIC, Valencia-CSIC, and D.F.A.M.N., U. de Valencia,
     Avda. Dr. Moliner 50, ES-46100 Burjassot (Valencia), Spain
    \label{VALENCIA}}
\titlefoot{Institut f\"ur Hochenergiephysik, \"Osterr. Akad.
     d. Wissensch., Nikolsdorfergasse 18, AT-1050 Vienna, Austria
    \label{VIENNA}}
\titlefoot{Inst. Nuclear Studies and University of Warsaw, Ul.
     Hoza 69, PL-00681 Warsaw, Poland
    \label{WARSZAWA}}
\titlefoot{Fachbereich Physik, University of Wuppertal, Postfach
     100 127, DE-42097 Wuppertal, Germany
    \label{WUPPERTAL}}
\addtolength{\textheight}{-10mm}
\addtolength{\footskip}{5mm}
\clearpage
\headsep 30.0pt
\end{titlepage}
%
\pagenumbering{arabic} 
\setcounter{footnote}{0} %
\large
%
\section{Introduction}

\begin{figure}[b]
\begin{center}
\mbox{\epsfxsize\textwidth\epsfbox[50 50 550 550]{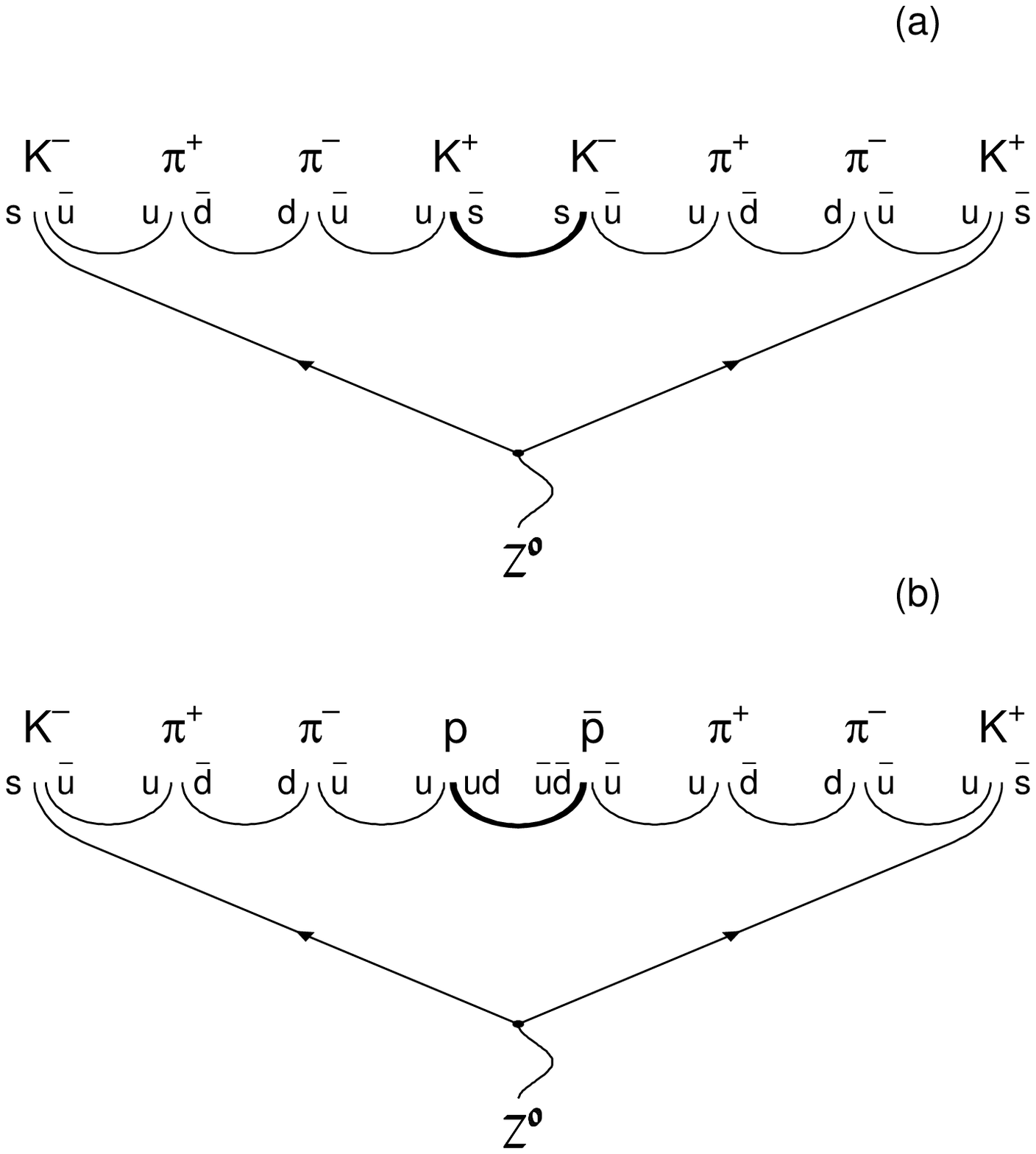}}
\end{center}
\caption[]{Illustration of \KK and \pp production in the string model.
Each loop represents a \qqb or diquark anti-diquark pair produced from
potential energy in the string.
(a) Production of \KK from an \ssb virtual pair. 
(b) Production of \pp from a diquark anti-diquark pair.}
\label{fig:pstring}
\end{figure}

The analysis of correlations between particles produced in hadronic \zz
decay is an effective tool for studying the fragmentation process.
In particular, tests can be made of two basic classes of
fragmentation models, `string' and `cluster' types, represented in this
study by Jetset 7.3 \cite{jetset} and Herwig 5.9 \cite{herwig}, respectively.
Distinct differences in predictions from these models occur for certain
particle pair correlations.
In particular, the orientation in rapidity and in \pT of particle pairs
is expected to be a distinguishing feature between models.
Charged particle pairs, adjacent or nearby in rapidity, are predicted to
be produced in a different way for the string and cluster models.
For \KK and \pp pairs the string model predicts a definite rank-ordering
in rapidity, with respect to the direction from primary quark to
antiquark.
Rapidity-rank is defined as the position a particle has in the rapidity
chain after ordering the particles in an event according to their rapidity
values.
Rapidity ordering is expected to correspond closely to string-rank
ordering (position on the string) as pictured in Figure~\ref{fig:pstring}.
By contrast, the cluster model produces \pp pairs, and \KK pairs (partially), 
via the isotropic decays of clusters.
The clusters are rank-ordered; however, their decay products are not
necessarily in rank-order.
Consequently, one expects differences in the predictions for correlations
in rapidity and \pT from the two models.

Asymmetric orientations of particle pairs in rapidity are expected from
string-fragmentation models.
In these models, mesons are formed from string elements when breaks occur
between virtual flavour-neutral \qqb pairs.
Baryons are considered to be formed when breaks occur between diquark
anti-diquark pairs.
An asymmetry occurs because each \qqb loop breaks such that the \qb is
always nearer on the string to the primary quark, and the \q nearer to the
primary antiquark.
For the diquark anti-diquark case, the converse orientation arises.
This process causes \KK and \pp pairs to assume an asymmetric rapidity
orientation, termed here rapidity-alignment, as shown in
Figure~\ref{fig:pstring}.
First indications of ordering along the quark-antiquark axis have been
reported by the SLD Collaboration \cite{SLD}.

The cluster model can be pictured approximately by replacing the hadrons in
Figure~\ref{fig:pstring} by clusters which decay isotropically, usually
into two hadrons.
In this model, the specific cluster mass spectrum and the assumption of
isotropic decay will affect particle-pair correlations in rapidity and in
transverse momentum.
In the following, the string and cluster models, with their different
hadronization mechanisms, are compared to the data.

\section{Data Sample and Event Selection}

This analysis is based on data collected with the DELPHI detector
\cite{delphi} at the CERN LEP collider in 1994 and 1995 at the \zz
centre-of-mass energy.
The charged particle tracking information relies on three
cylindrical tracking detectors (Inner Detector, Time Projection
Chamber (TPC), and Outer Detector) all operating in a \mbox{1.2 T}
magnetic field.

The selection criteria for charged particles are: momentum
above 0.3 GeV/c, polar angle between $15\de$ and $165\de$ and
track length above 30 cm.
In addition the impact parameters with respect to the interaction point
are required to be below 0.5 cm perpendicular to and 2.0 cm along the
beam.
These impact parameter cuts decrease the number of protons which
result from secondary interactions in the detector. 
Also, protons from $\Lambda$ and $\Sigma$ decays are largely removed.

Hadronic \zz decays are selected by requiring at least three charged
particle tracks in each event hemisphere, defined by the `thrust' axis
(see next section), and a total energy of all charged particles exceeding
15 GeV.
The number of hadronic events is $\sim$ 2 million.

Charged particle identification is provided by a tagging procedure
which combines Cherenkov angle measurement from the RICH detector
with ionization energy loss measured in the TPC.
Details on the particle identification can be found in reference
\cite{delphi}.
In addition, the polar angle for identified particles is restricted
to be in the barrel region, between $47\de$ and $133\de$.

\section{\bfm Rapidity-Alignment of \KK and \pp Pairs}

In the following, the rapidity-alignment resulting from an asymmetric
orientation of \KK or \pp pairs in rapidity with respect to the primary
quark-to-antiquark direction is investigated.
The rapidity, $y$, is defined as
${1\over2} \, {{ln}}\bigl((E+p_L) / (E-p_L) \bigr)$, where $p_L$ is
the component of momentum parallel to the thrust axis, and $E$ is
the energy calculated using the RICH determined particle mass.
The thrust approximates the directions of the primary $q$ and
$\bar{q}$, especially for two-jet events.
Specifically, a study is made of the preference for either the positive,
or negative, charged member of the pair to be nearer in rapidity to the
particle containing the primary quark rather than primary antiquark,
see Figure~\ref{fig:pstring}.

To study the rapidity-alignment of particle pairs it is necessary to 
determine the directions of the primary quark and antiquark.
Advantage is taken of the fact that a primary \ssb initiated event will
frequently hadronize yielding a high momentum \Kp and \Kmx.
An effective tagging of the directions of the primary quark (\sxx) and the
antiquark (\sxbx) is achieved by selecting events where the highest
momentum particle in each hemisphere, defined by `thrust', is a charged
kaon (a \Km in one hemisphere, and a \Kp in the other). 

The data selected for this study are those tagged events (i.e., where the
leading particle in each hemisphere is a kaon) that contain either an
additional \KK pair ({\it not} the tagged pair), or a \pp pair.
The restriction is made that events have only `one \Kp and one \Kmx'
({\it not} including the tagged $K$'s), or only `one \p and one \bpx \,'
in a given hemisphere.
Hemispheres are defined, one for positive $y$ and one for negative
$y$, with respect to the thrust direction.
Each hemisphere is considered independently.
Essentially no events had additional \KK or \pp pairs in both hemispheres.
For the \KK events the combined-probability tag for particle identification 
is required to be at the `standard' level, see Ref \cite{delphi}.
Because of reduced statistics, the combined-probability tag for \pp events
was taken at the `loose' level.
In each case, these respective levels also apply to the primary quark kaon
tags.
This selection yields 1250 events for \KKx, and 835 events for \ppx.

The same procedure was used on Monte Carlo events from Jetset.
Standard DELPHI detector simulation along with charged particle
reconstruction and hadronic event selection are applied to the events from
Jetset with parameters tuned as in reference \cite{tune}.
The number of selected Monte Carlo events for \KK (\pp) was found to be 
1.02 (1.27) times that of the data, for equal luminosity.
The difference in \pp between data and Monte Carlo may result from some
deficiency in the fragmentation properties of the model.

A study of events from Jetset including detector simulation determined
the purity for the \KK tagged events with an additional \KK or \pp pair 
to be $\sim$ 45\% and $\sim$ 35\%, respectively.
The event detection efficiency for events with \KK or \pp pairs is
$\sim$ 7\% and $\sim$ 12\%, respectively, resulting from the
requirement to identify four particles (K's or p's) in each event.
These values are nearly constant over the range of the analysis variable
\dy defined later.
The purity is computed from the ratio of Jetset events detected and
congruous with a generated event, to the total number of events detected.
The efficiency is obtained from the ratio of Jetset events detected to 
the total number of events generated.

\begin{figure}[b]
\begin{center}
\mbox{\epsfxsize\textwidth\epsfbox[-50 0 550 550]{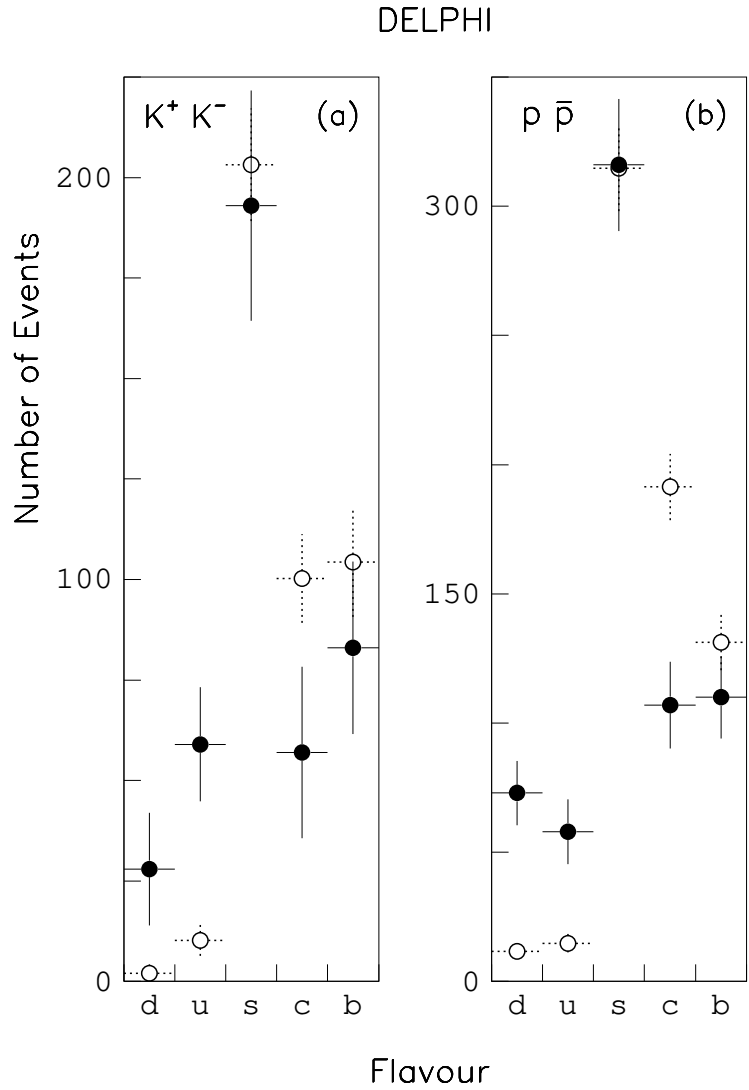}}
\end{center}
\caption[]{Distribution of primary-quark flavour from Jetset and Herwig
shown by the solid and open circles, respectively, for events that have
been tagged by high momentum charged $K$'s. 
(a) For \KK production (in addition to the tagged $K$'s).
(b) For \pp production.}
\label{fig:flavour}
\end{figure}

The distributions of primary-quark flavour, for the events that are
tagged, are shown in Figure~\ref{fig:flavour} for Jetset and Herwig by
the solid and open circles, respectively.
A background subtraction of like-sign pairs has been applied to account
for uncorrelated kaon or baryon pairs.
As seen, the \ssb contribution is enhanced for both the \KK and \pp events.
It is also possible for \ccb primary quarks to generate rapidity
alignment.
This results from the production of \Dz[\Dp] mesons from the $c$ quark,
which strongly favour decays to \Km rather than \Kp; the opposite occurs
for the $\bar{c}$.
The \bbb primary quarks also produce an effect though through a 
longer chain from $B$ meson to $D$ meson to kaon.

The operational definition of rapidity-alignment is given as follows.
For a \KK pair to be {\it in} rapidity-rank order, the \Kp of the pair
should be nearer in the rapidity chain to the tagged \Kmx.
Equivalently, the \Km from the pair should be nearer to the tagged \Kpx.
This is depicted in Figure~\ref{fig:pstring}, assuming a correspondence
between string-rank and rapidity-rank order.
Similarly, for a \pp pair the \p of the pair should be nearer in rapidity
to the tagged \Km; and the \bp from the pair nearer to the tagged \Kpx.

Since particle pairs with small rapidity difference between them have a
high probability to have `crossed-over' (reversed rank), this study is
performed as a function of the absolute rapidity difference between the
two particles of the pair, ${\mit\Delta} y = | y^+ - y^- |$, where $y^+$
($y^-$) are the rapidities of the positive (negative) members of the pair.
Rapidity-rank cross-overs can result from resonance/cluster decays, hard
gluon production and \pT effects.

\begin{figure}[b]
\begin{center}
\mbox{\epsfxsize\textwidth\epsfbox[-25 0 550 550]{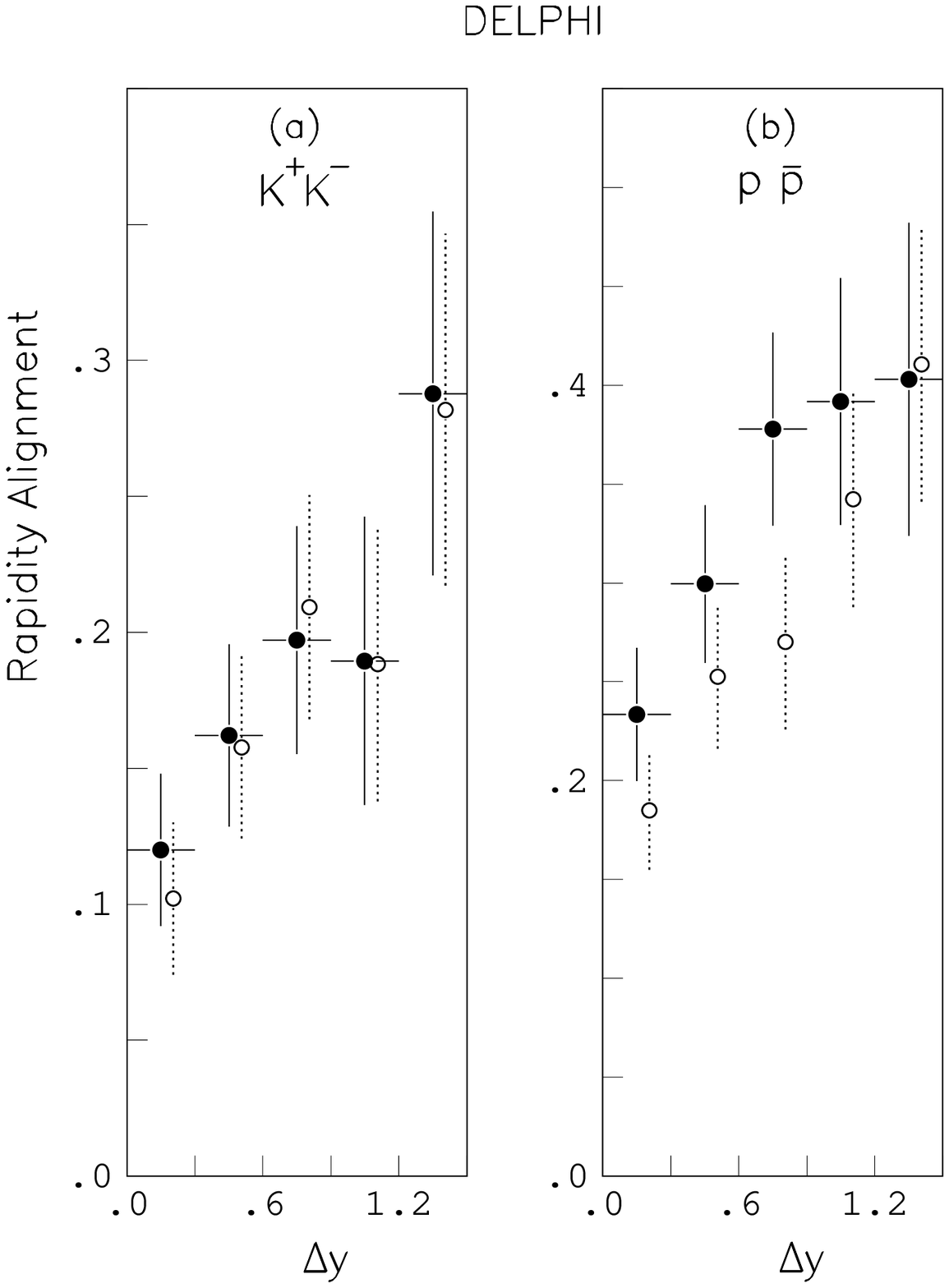}}
\end{center}
\caption[]{Rapidity-alignment, uncorrected, as a function of the
absolute rapidity difference, \dyx, between the particle pair.
The data points are indicated by the solid circles, and the predictions 
of Jetset are shown by the open circles.
The plot is an integral; i.e., all pairs with \dy greater than a
given (abscissa) value are plotted.
(a) for \KK pairs. (b) for \pp pairs.}
\label{fig:polar1}
\end{figure}

\begin{figure}[b]
\begin{center}
\mbox{\epsfxsize\textwidth\epsfbox[-25 0 550 550]{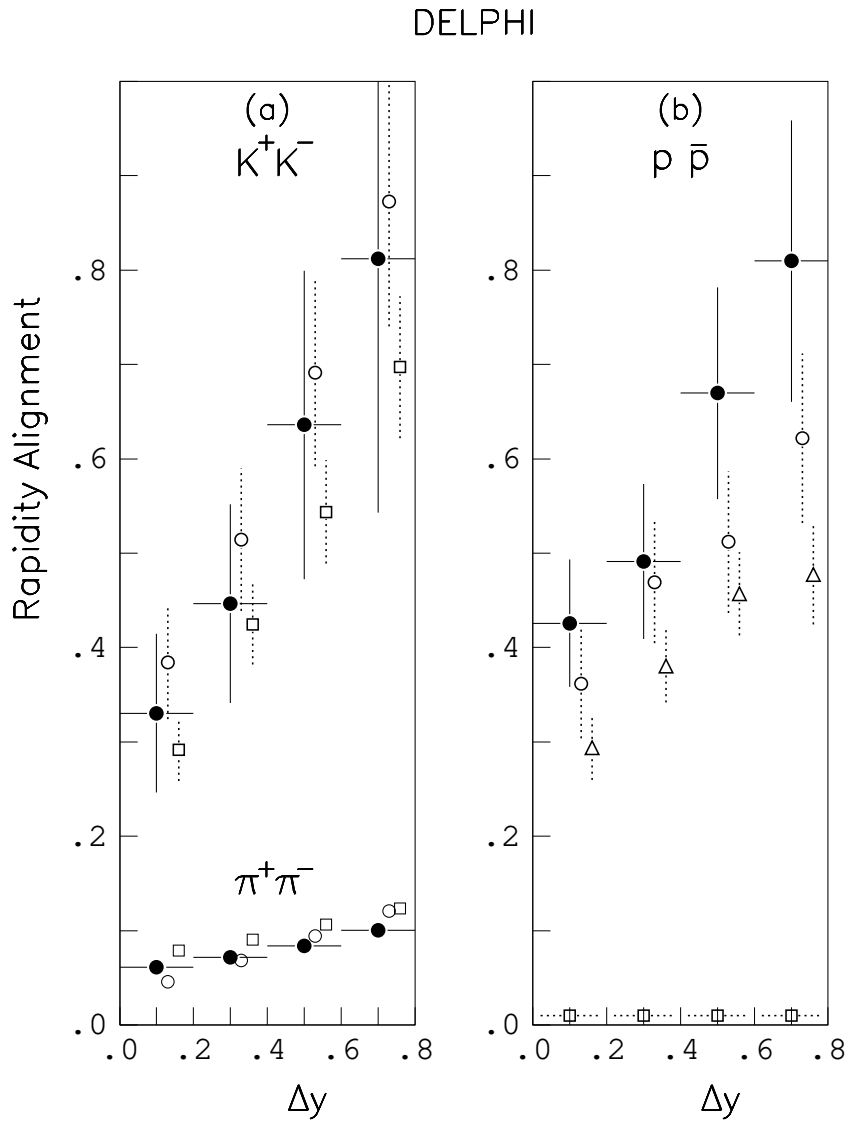}}
\end{center}
\caption[]{Rapidity-alignment, corrected for background, as a function
of the absolute rapidity difference, \dyx, between the particle pair.
The data points are indicated by the solid circles, and the predictions of
Jetset and standard Herwig are shown by the open circles and squares,
respectively.
The plot is an integral; i.e., all pairs with \dy greater than a
given (abscissa) value are plotted.
(a) for \KK pairs and \pipi pairs;
the \pipi pairs do not have like-sign pair subtraction, see text.
(b) for \pp pairs.
The Herwig result with the process gluon to diquarks included is shown
by the open triangles.}
\label{fig:polar2}
\end{figure}

The rapidity-alignment variable, $\cal P$, for \KK and for \pp pairs is
defined as follows,
              $${\cal P}\bigl(\dyx\bigr)=
    \Bigl(\Ninx-\Noutx\Bigr)\Big/\Bigl(\Ninx+\Noutx\Bigr),  \eqno(1)$$
where \Nin and \Nout are defined as the number of particle pairs with their
charges {\it in} and {\it out} of rapidity-rank order, and are implicitly 
a function of \dyx.
In Figures~\ref{fig:polar1} and \ref{fig:polar2}, the calculation is
given as an integral over \dyx; that is, all pairs with \dy greater than
a given (abscissa) value are plotted.
In the absence of rapidity-alignment \Nin and \Nout should be equal,
within statistics, for all \dyx.

The uncorrected values of \PPPdy for \KK and \pp pairs are displayed
in Figures~\ref{fig:polar1}(a) and (b), respectively.
The data, shown by the solid circles, for both \KK and \pp exhibit a
definite rapidity-alignment which increases with \dyx.
The predictions from Jetset, shown by the open circles, are in good
agreement with the data.
Herwig predictions incorporating detector simulation with
misidentifications were not available for a comparison to the uncorrected
values.

A correction has been applied to the values of \PPPdy for the data to
account for particle misidentifications and uncorrelated kaon or baryon
pairs.
This is achieved by a background subtraction of like-sign kaon or
baryon pairs. 
The like-sign pairs provide a direct measure of the background which
would be contained in the \KK and \pp pairs.
Since misidentifications or uncorrelated pairs would not favour a rank
ordering, the background represented by one half of the like-sign pairs is
subtracted from both {\it in} and {\it out} of order pairs equally.
The corrected values of \PPPdy for \KK and \pp pairs are displayed
in Figures~\ref{fig:polar2}(a) and (b), respectively.
The data, shown by the solid circles, now display a stronger
rapidity-alignment with values approaching 1.0 for large \dyx.
However, larger errors result from the subtraction procedure.

For the study that incorporates corrections for misidentifications, the
model predictions from Jetset and Herwig are taken from the generation
level with momentum and angle cuts as were applied to the data.
For Jetset the predictions are in agreement with those obtained from the
detector simulation version which allows for misidentifications.
The predictions from the string-model (Jetset) shown by the open circles
are in agreement with both the \KK and \pp data.
The cluster-model (Herwig) predictions, shown by open squares, also 
give agreement with the \KK data.
However, Herwig predicts no alignment for the \pp case.
This occurs because baryon anti-baryon pairs are produced jointly from 
individual isotropically-decaying clusters, see Ref \cite{herwig}.
The `detected' \pp pairs cannot be rapidity aligned, whether they
originate from the same cluster or from different clusters.
It was found that 70\% of detected \pp pairs come from the same cluster.

The Herwig program is customarily employed in the `standard' mode that does
{\it not} include the non-perturbative process of splitting gluons into 
diquark anti-diquark pairs, see Ref \cite{herwig}.
If the gluon-to-diquarks process were allowed, rapidity-aligned \pp
pairs could be produced.
The diquark and anti-diquark would separately adjoin into adjacent clusters
which are rank ordered.
In this case, a proton produced from the diquark, and an anti-proton
produced from the anti-diquark would be rank ordered since they do not
originate from the same cluster as in the case of standard Herwig.
The rank ordering then allows the pairs to be rapidity-aligned.
Herwig has a provision for this process, which is controlled by a scale
and a rate parameter.
The scale parameter sets the maximum four-quark mass sum, see below; and
the rate parameter is a measure of the amplitude for the process.
These parameters are normally set to the default values which suppress the
non-perturbative process; the default rate parameter for this process is
zero.

To attempt to reproduce the data with Herwig, the scale parameter for the
gluon-to-diquarks process was set to 1.9, to include light diquarks (below 
the four ({\it s}) threshold).
The rate parameter was set to the value 10.0,
which produces a sufficient rapidity-alignment.
The prediction, for these parameter values, is shown by the open triangles
in Figure~\ref{fig:polar2}(b).
The agreement with the data in this case is satisfactory.
This new parameter setting, however, causes the predicted baryon multiplicity to
increase, and to be inconsistent with the data.
This problem can be mitigated by reducing the value of the a priori
weight-parameter, for the splitting of clusters to baryons, to
$\sim$ 0.25, from the tuned value of 0.74, Ref \cite{tune}.
The a priori hypothesized default value is 1.0.

For completeness, the potential rapidity-alignment of \pipi pairs is 
also examined.
In this case, the large pion multiplicity with a string-like alternating
charge structure should preclude \pipi pairs from having any substantial 
rapidity-alignment, see Ref \cite{charge}.
Each \pipi combination in a given event is included in the calculation of
\PPPdy.
Predominant pion production with normal occurrence of like-sign pion pairs
makes background subtraction incongruous in this case.
The subtraction technique used for \KK and \pp pairs to correct for 
particle misidentifications is not necessary for like-sign pion pairs
since misidentifications are not their primary source.
For charged pion identification the combined-probability tag is set 
at the `standard' level.
The values of \PPPdy for the \pipi data, shown by the solid circles, are
included in Figure~\ref{fig:polar2}(a).
The rapidity-alignment is quite small, as compared to that from \KK
and \ppx.
The small, but finite, rapidity-alignment indicates an incomplete 
cancellation of {\it in} and {\it out} of order \pipi pairs expected from
a string-like alternating charge structure.
The predictions from Jetset and Herwig, shown by the open circles and open
squares, respectively, are in qualitative agreement with the data.

\section{\bfm $\mrm{p}_T$ Compensation of \pipix, \KK and \pp Pairs}

The mechanism of local \pT conservation (i.e., compensation) can be studied 
from correlations in \pT between particles which are adjacent or nearby 
in rapidity.
The azimuthal-angle difference between particles, \dphix, measured in the
\pTx\,-plane, where \pT is transverse to the thrust direction, is the
variable used in this analysis.
The tagging procedure used for the rapidity-alignment analysis is
not employed here.

According to the string and cluster models, \pT correlations should be stronger
for oppositely-charged like-particle pairs.
In the case of the string model, this can be understood from examination of
Figure~\ref{fig:pstring}, where adjacent hadrons share quarks from a 
single `breakup' vertex.
For cluster models, \pT correlations can also be expected, for example, from
the fact that clusters frequently decay into like-particle pairs.

The particle pairs \pipix, \KKx, and \pp, determined with the particle
identification combined-probability tag at the `standard' level,
are used in this study \cite{delphi}.
The requirement is made that the absolute relative difference in \pT
between the two particles,
$|p_{T}^{+}-p_{T}^{-}|/.5(p_{T}^{+}+p_{T}^{-})$, be less than 0.1.
In this formula \pT is treated as a scalar.
In addition, the thrust value for the event is required to be greater than
0.95 for the events with \pipi and \KK pairs, and, because of reduced
statistics, greater than 0.9 for the \pp case.
These stringent conditions provide better definition of \pT with respect
to the initial \qqb direction, and thus more sensitivity to differences
between the string and cluster models.
Also, for the \pipi pairs, the particles are required to be adjacent in
rapidity (rapidity-ranks differ by one unit).
For the \KK pairs, the rapidity-ranks are allowed to differ by two units
in order to increase statistics.
For the \pp pairs, this condition is not applied because of reduced statistics.

\begin{figure}[b]
\begin{center}
\mbox{\epsfxsize\textwidth\epsfbox[0 0 550 550]{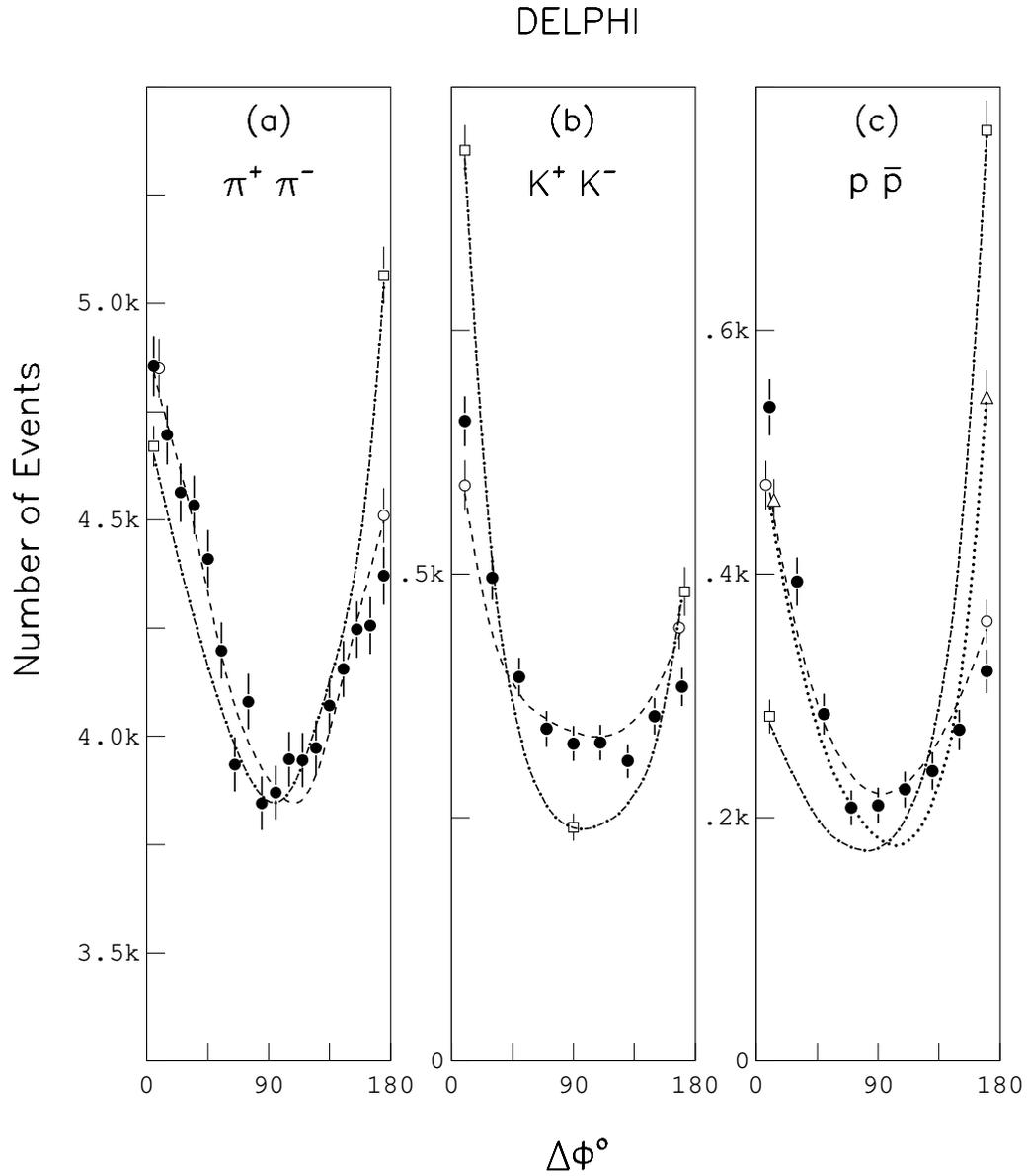}}
\end{center}
\caption[]{Azimuthal-angle difference, \dphix, between particles from the
oppositely-charged particle pairs \pipix, \KKx, and \pp.
\dphi is measured in the \pTx\,-plane, where \pT is transverse to the 
thrust direction (defined in text).
In addition, a requirement on thrust and \pT is made (see text).
The data points are indicated by the solid circles.
The predictions from Jetset and standard Herwig are shown by the dashed 
and dot-dashed lines, respectively.
Error bars for Jetset (open circles) and Herwig (open squares) are shown
for selected points (see text).
(a) for \pipi pairs, adjacent in rapidity.
Plot has suppressed zero.
(b) for \KK pairs, rapidity-ranks differ up to two units.
(c) for \pp pairs, without a rapidity condition.
The prediction from Herwig with the process gluon-to-diquarks implemented
is shown by the dotted line, with error bars and open triangles for
selected points.}
\label{fig:dphi}
\end{figure}

The uncorrected distributions of \dphi are shown in
Figures~\ref{fig:dphi}(a), (b), and (c) for \pipix, \KKx, and \pp pairs,
respectively.
The data are represented by the solid circles.
The string and cluster model predictions, from Jetset and standard Herwig,
are shown by the `dashed' and `dot-dashed' lines, respectively.
Error bars for Jetset (open circles) and Herwig (open squares) are shown
for selected points.
The Jetset errors are statistical.
Since Herwig did not give good agreement with the data, systematic
errors were evaluated by varying the parameters from Herwig according
to the fit results of reference \cite{tune}.
Only the cluster-mass cutoff parameter had a significant effect on the
\dphi distribution.

In all three cases, the Jetset predictions give significantly better
agreement with the data than does Herwig.
However, Jetset does not predict quite enough peaking at $\dphix=0\de$
for the \KK and \pp pairs.
It should be mentioned that although Jetset contains \pT conservation 
in each string breakup, it does not take into account local \pT
compensation between neighbouring vertices, see Ref \cite{local}.

A clear disagreement occurs for the model Herwig which predicts a stronger 
peaking of the distribution at $\dphix=180\de$, for all three cases, as
compared to the data.
This comes about because the clusters decay predominantly into two
particles which emerge back-to-back in \pTx, for clusters with small
initial \pTx.
This `limitation' of the Herwig model design arises from the assumption of
isotropically-decaying clusters which is basic to the model.

For comparison to the \pp data one might expect better agreement from the
Herwig model which incorporates the gluon-to-diquarks process.
The prediction, for this case, is shown by the `dotted' line in
Figure~\ref{fig:dphi}(c), with error bars and open triangles for 
selected points.
As seen there is significant improvement in the Herwig prediction, 
however it still fails to give an adequate description.

\section{Conclusions}

The rapidity-alignment of \KK and \pp pairs resulting from their
asymmetric orientation in rapidity, with respect to the direction from primary
quark to antiquark, is observed.
The Jetset string model agrees well with both the \KK and \pp data.
The Herwig cluster model agrees with the \KK data; however, agreement for
the \pp case can be achieved only if the non-perturbative process
gluon-to-diquarks is implemented.
In this case, the predicted baryon production is greatly increased.
Inauspiciously, in order to obtain agreement with the data, it is essential 
to reduce the a priori weight-parameter (hypothesized value 1.0) for
splitting a cluster to form baryons to $\sim 0.25$.

Particle pair correlations in \pT are in much better agreement with Jetset
than with Herwig.
Herwig fails to describe the \pT correlations for all cases, \pipix,
\KKx, and \pp because of the mechanism of isotropically-decaying clusters
inherent in the model.

\subsection*{Acknowledgements}
\vskip 3 mm
 We are greatly indebted to our technical 
collaborators, to the members of the CERN-SL Division for the excellent 
performance of the LEP collider, and to the funding agencies for their
support in building and operating the DELPHI detector.\\
We acknowledge in particular the support of \\
Austrian Federal Ministry of Science and Traffics, GZ 616.364/2-III/2a/98, \\
FNRS--FWO, Belgium,  \\
FINEP, CNPq, CAPES, FUJB and FAPERJ, Brazil, \\
Czech Ministry of Industry and Trade, GA CR 202/99/1362,\\
Commission of the European Communities (DG XII), \\
Direction des Sciences de la Mati$\grave{\mbox{\rm e}}$re, CEA, France, \\
Bundesministerium f$\ddot{\mbox{\rm u}}$r Bildung, Wissenschaft, Forschung 
und Technologie, Germany,\\
General Secretariat for Research and Technology, Greece, \\
National Science Foundation (NWO) and Foundation for Research on Matter (FOM),
The Netherlands, \\
Norwegian Research Council,  \\
State Committee for Scientific Research, Poland, 2P03B06015, 2P03B1116 and
SPUB/P03/178/98, \\
JNICT--Junta Nacional de Investiga\c{c}\~{a}o Cient\'{\i}fica 
e Tecnol$\acute{\mbox{\rm o}}$gica, Portugal, \\
Vedecka grantova agentura MS SR, Slovakia, Nr. 95/5195/134, \\
Ministry of Science and Technology of the Republic of Slovenia, \\
CICYT, Spain, AEN96--1661 and AEN96-1681,  \\
The Swedish Natural Science Research Council,      \\
Particle Physics and Astronomy Research Council, UK, \\
Department of Energy, USA, DE--FG02--94ER40817. \\

\clearpage

\clearpage

\clearpage

\clearpage

\clearpage

\end{document}